\documentstyle[12pt]{article}

\begin{document}

\pagestyle{plain}

\begin{center}
{\bf FEYNMAN\ INTEGRAL\ APPROACH\ TO\ ABSORPTION\ IN\ QUANTUM\ MECHANICS }

\vspace{1cm} A. Marchewka\\[5mm]
Department of Physics\\[0pt]
Tel-Aviv University\\[0pt]
Ramat-Aviv, Tel-Aviv 69978, Israel\\
e-mail marhavka@ccsg.tau.ac.il\\[5mm]
Z. Schuss\\[5mm]
Department of Mathematics\\[0pt]
Tel-Aviv University\\[0pt]
Ramat-Aviv, Tel-Aviv 69978, Israel\\
e-mail schuss@math.tau.ac.il

\vspace{1cm} \vspace{3mm} {\bf ABSTRACT}
\end{center}

\vspace{5mm} We propose a formulation of an absorbing boundary for a
quantum particle.  The formulation is based on a Feynman-type integral 
over trajectories that are confined by the absorbing boundary.   
Trajectories that reach the absorbing wall are instantaneously
terminated and their probability is discounted from the population 
of the surviving trajectories. This gives rise to a unidirectional
absorption current at the boundary. We calculate the survival
probability as a function of time. Several modes of absorption are
derived from our formalism: total absorption, absorption that depends
on energy levels, and absorption of non-interacting particles. Several
applications are given: the slit experiment with an absorbing screen and
with absorbing lateral walls, and one dimensional particle between two
absorbing walls. The survival probability of a particle between
absorbing walls exhibits decay with beats.
\newpage 

\pagestyle{plain} 
\setcounter{page}{1}

\noindent
{\large {\bf 1. Introduction}}\newline
\renewcommand{\theequation}{1.\arabic{equation}} \setcounter{equation}{0}

There are many different descriptions of absorption in electromagnetic
theory, nuclear physics, solid state physics, and so on. These descriptions
are mostly based on scattering theory. In scattering theory absorption
occurs when a jet of particles hits an obstacle and is partially absorbed.
The reduction in the population of particles is described in scattering
theory by a complex potential. The introduction of a complex potential in
scattering theory is supported by and large by phenomenological
considerations \cite{Hodgson}.

The introduction of a complex potential in scattering theory, that is, of
phase shift, is analogous to the introduction of a complex dielectric
coefficient in Maxwell's equations. There exist several models for the
calculation of the complex part of the dielectric constant in Maxwell's
theory. The basis for these models is the possibility to write equations of
motion of a damped charged particle in an electromagnetic field and to
identify the complex part of the dielectric constant as the damping
coefficient \cite{Landau,Jac}.  Another approach to decay is based on the 
decoherent histories approach \cite{Halliwell}.

The purpose of this paper is to propose a Feynman-type integral to describe
total or partial absorption of particles in a surface bounding a domain. 
The need for such a description stems from the reflecting property of
potentials of all kinds, real and complex, finite and infinite. This
property eliminates potentials as means of absorbing all particles 
that reach a given absorbing surface such as a photographic plate.
Additional postulates are needed to incorporate absorbing surfaces into
quantum mechanics. The method of complex potentials, as used in
Maxwell's equations, does not carry over to quantum mechanics due to 
the fact that the wave function of a particle does not interact with 
the medium the way an electromagnetic wave does. In particular, 
in classical quantum theory, unlike in electromagnetic theory, the 
wave function does not transfer energy to the medium.

In our approach to absorption, we postulate an absorbing boundary that
absorbs instantaneously all Feynman trajectories at the instant they
reach the boundary. The probability of these trajectories is
discounted from the wave function of the surviving trajectories, much
like the procedure adopted in the derivation of Fermi's Golden Rule. 
We find that the wave function of the surviving Feynman trajectories satisfies
Schr\"odinger's equation with reflecting boundary conditions on the
absorbing boundary. We define a discounted wave function whose total
probability within the absorbing boundaries is the survival
probability of the  un-absorbed trajectories. We find the decay law of
the survival probability for different modes of absorption: total
absorption, absorption that depends on energy levels, and absorption 
of non-interacting particles. Total absorption is obtained if the wave
function is discounted by the probability of the absorbed
trajectories. In this mode of absorption the survival probability
decays in time at a rate proportional to the total energy of the system with
beats at Bohr frequencies. Absorption that depends on energy levels apportions
different decay rates to modes corresponding to different energies in
a such a away that the total discounted probability is the same as in the
former case. Absorption of non-interacting particles produces a
discounted wave function which is the product of the discounted wave
functions of each particle separately.

In Section 2, we formulate the postulates of absorption in a surface
and explain the discounting procedure. In Section 3, we calculate the
Feynman integral over the class of trajectories that are bounded
within a given boundary and show that it satisfies Schr\"odinger's
equation with total reflection on the boundary. In Section 4, we
calculate the probability of Feynman trajectories that propagate into
the absorbing boundary and obtain the survival probability of the
un-absorbed trajectories. In Section 5, we describe other modes of
absorption. In Section 6, we calculate the absorption current at the
absorbing boundary and consider the slit experiment with an absorbing
screen. Section 7 contains the following examples: a particle between
absorbing walls, a Gaussian wave packet
incident on an absorbing wall, and the slit experiment with 
absorbing lateral boundaries. Experiments are proposed to examine the
various decay modes. Finally, a discussion and summary are offered in
Section 8.\\

\noindent
{\large {\bf 2. The postulates of absorption in a surface}}\newline
\renewcommand{\theequation}{2.\arabic{equation}} \setcounter{equation}{0}

The two simplest types of instantaneous absorption at a wall are the
absorption of all Feynman trajectories when they reach the wall for the
first time, or absorption of trajectories that propagate across the wall. It
was shown in \cite{PLA} that Feynman integrals over the set of trajectories
that terminate at a given wall produce wave functions that vanish on and
beyond the wall. The resulting wave functions are continuous, though have
discontinuous derivatives on the boundaries of their supports. It was shown
in \cite{MSU} that for this type of initial wave functions the
probability that propagates across the boundary of the support in a short
time $\Delta t$ is proportional to $\Delta t^{3/2}$ so that a continuous
discounting of the wave function by the probability of the trajectories that
crossed into the absorbing domain beyond the wall leads to total reflection,
much like in the Zeno effect \cite{Peres}. It is shown below that the
probability density that propagates into an absorbing wall in a short time 
$\Delta t$ is proportional to $\Delta t$, so that discounting  the
probability of these trajectories leads to a decay law, consequently, we
adopt the concept of instantaneous absorption at a wall for wave functions
that vanish on and beyond the wall. 

More specifically, in our Feynman-type integral trajectories that propagate
into the surface for the first time are considered to be absorbed
instantaneously or reflected with a given probability which is a property of
the interaction between the particle and the absorbing surface. The absorbed
trajectories are therefore terminated at that surface. The population of the
surviving trajectories is discounted by the probability of the absorbed
trajectories at each time step. The instantaneous absorption rate is assumed
proportional to the number (probability) of particles at the absorbing
surface at a given moment of time. The proportionality constant is a
characteristic length, $\lambda $, determined by the absorbing material, 
{\em e.g.}, photographic plate or fluorescent screen, and by the absorbed
particle, {\em e.g.}, electron or neutron. The length $\lambda $ is assumed
to depend weakly on the absorbed particle's energy so, in this paper, we
approximate it by a constant. Under this assumption $\lambda $ can be
determined experimentally by measuring the absorption rate of a single
energy level and than be used to calculate the absorption rate at any other
energy level, in some energy range, or the simultaneous absorption of
several levels in this range. The constant $\lambda $ can be assumed to
contain a multiplicative factor that represents the probability of
absorption of a trajectory when it hits the wall.

In the context of the decay of an energy level into a continuum, usually
described by the Wigner-Weiskopf theory and Fermi's Golden Rule \cite{CT},
the characteristic length $\lambda$ in our theory, can be related to the
size of the support of the perturbing Hamiltonian. In the context of $\alpha 
$-decay in nuclear physics, the length $\lambda $ can be related to the
width of the potential barrier \cite{Nuclear}.

A detailed theory of $\lambda $ can be expected to involve the atomic detail
of the absorbing wall and the details of the interaction between the
particle and the atoms of the absorber. In contrast, our theory involves the
trajectories of the particles before they hit the absorbing wall so that $%
\lambda $ is an external input into our theory. This situation is analogous
to the theory of frequency dependent refraction coefficient in
electromagnetic theory where the frequency dependence is determined from a
detailed analysis of the interaction between the electromagnetic wave and
the atoms of the medium. Thus the frequency dependent refraction coefficient
is an external input into the Helmholtz equation. In this theory, the
knowledge of the response to a single frequency does not determine in a
simple way the response to a wave containing several frequencies. Only in
the case of a frequency independent dielectric constant its measurement at
one frequency allows to predict the response to a wave containing several
frequencies. In different theories of absorption the parameter $\lambda $
can depend on any number of physical properties of the system, such as the
energy of the system or some average energy per mode, and so on.

In general, if the trajectories are partitioned into two subsets, the part
of the wave function obtained from the Feynman integral over one subset
cannot be used to calculate the probability of this subset, due to
interference between the wave functions of the two subsets. However, in the
physical situation under consideration, such a calculation may be justified
as follows. Our procedure, in effect, assumes a partition of all the
possible trajectories at any given time interval $[t,t+$ $\Delta t]$ into
two classes. One is a class of {\em restricted trajectories} that have not
reached the absorbing surface by time $t+\Delta t$ and remain in the domain,
and the other is a class of trajectories that arrived at the surface for the
first time in the interval $[t,t+$ $\Delta t]$. We assume that the part of
the wave function obtained from the Feynman integral over trajectories that
reached the surface in this time interval no longer interferes with the part
of the wave function obtained from the Feynman integral over the class of
restricted trajectories in a significant way. That is, the interference is
terminated at this point so that the general population of trajectories can
be discounted by the probability of the terminated trajectories. This
assumption makes it possible to calculate separately the probability of the
absorbed trajectories in the time interval $[t,t+$ $\Delta t]$.\newline

The assumptions discussed above can be summarized in the form of the
following postulates:\newline

\noindent
I. {\em The absorption of Feynman trajectories represents absorption of
actual trajectories of particles.}\newline

\noindent
II. {\em The population of Feynman trajectories can be discounted by the
probability of the absorbed trajectories.}\newline

\noindent {\large {\bf 3. The Feynman integral over bounded trajectories}}\newline
\renewcommand{\theequation}{3.\arabic{equation}} \setcounter{equation}{0}

If \ a quantum particle is constrained in space to a finite (or semi finite)
domain, the Feynman integral has to be confined to Feynman trajectories that
stay forever in this domain. This implies the following modification in the
definition of the Feynman integral. The function space is now the class 
\[
\sigma _{a,b}=\left\{ x\left( \cdot \right) \in C\left[ 0,t\right]
\,|\,a\leq x\left( \tau \right) \leq b,\,0\leq \tau \leq t\right\} 
\]
and the definition of the Feynman integral over the class $\sigma _{a,b}$ is 
\begin{eqnarray}
K_{a,b}(x,t) &=&\int_{\sigma _{a,b}}\exp \left\{ \frac i\hbar S\left[
x(\cdot ),t\right] \right\} {\cal D}x(\cdot )  \nonumber \\
&&\mbox{}  \nonumber \\
&\equiv &\lim_{N\rightarrow \infty }\alpha _N\int_a^b\dots \int_a^b\exp
\left\{ \frac i\hbar {\cal S}(x_0,\dots ,x_N,t)\right\} \prod_{j=1}^{N-1}\,dx_j.
\label{FIab}
\end{eqnarray}
Next, following the method of \cite{Keller}, we show that $K_{a,b}(x,t)$
satisfies the Schr\"{o}dinger equation and determine the boundary conditions
at the endpoints of the interval $[a,b]$. We begin with a derivation of a
recursion relation that defines $K(x,t)$. We set 
\begin{equation}
K_N(x_N,t)\equiv \alpha _N\int_a^b\dots \int_a^b\exp \left\{ \frac i\hbar
{\cal S}(x_0,\dots ,x_N,t)\right\} \prod_{j=1}^{N-1}\,dx_j,  \label{Knab}
\end{equation}
then by the definition (\ref{FIab}), $K(x,t)=\lim_{N\rightarrow \infty
}K_N(x,t)$. The definition (\ref{Knab}) implies the recursion relation 
\begin{eqnarray}
\ &&K_N(x,t)=  \label{recursion} \\
&&\ \ \ \ \left\{ \frac m{2\pi i\hbar \Delta t}\right\} ^{1/2}\int_a^b\exp
\left\{ \frac i\hbar \left[ \frac{m(x-x_{N-1})^2}{2\Delta t}-V(x)\Delta
t\right] \right\} K_{N-1}(x_{N-1},t_{N-1})\,dx_{N-1}.  \nonumber
\end{eqnarray}
The following derivation is formal. A strict derivation can be constructed
along the lines of \cite{Keller}. We expand the function $%
K_{N-1}(x_{N-1},t_{N-1})$ in (\ref{recursion}) in Taylor's series about $x$
to obtain \newpage
\begin{eqnarray}
&&K_N(x,t)=  \label{Taylor} \\
&&\mbox{}  \nonumber \\
&&\left\{ \frac m{2\pi i\hbar \Delta t}\right\} ^{1/2}e^{-iV(x)\Delta
t/\hbar }\int_a^b\exp \left\{ \frac{im}{2\hbar \Delta t}(x-x_{N-1})^2\right\}
\nonumber \\
&&\mbox{}  \nonumber \\
&&{\bigg[}K_{N-1}(x,t_{N-1})-(x-x_{N-1})\frac{\partial K_{N-1}(x,t_{N-1})}{%
\partial x}+  \nonumber \\
&&\mbox{}  \nonumber \\
&&\frac 12(x-x_{N-1})^2\frac{\partial ^2K_{N-1}(x,t_{N-1})}{\partial x^2}%
+O\left( \left( x-x_{N-1}\right) ^3\right) \bigg]\,dx_{N-1}.  \nonumber
\end{eqnarray}
The integrals in eq.(\ref{Taylor}) are evaluated separately for $x$ inside
the interval $[a,b]$ and on its boundaries. When $x$ is inside the interval,
the integrals become the Fresnel integrals over the entire line in the limit 
$\Delta t\to0$. This recovers the Schr\"odinger equation inside the
interval. When $x=a,b$, the first integral in eq.(\ref{Taylor}) becomes in
the limit $\Delta t\rightarrow 0$ the Fresnel integral over half the real
line while the other integrals vanish. We obtain 
\begin{eqnarray}
K(a,t)={\frac 12}K(a,t),\label{half}
\end{eqnarray}
hence $K(a,t)=0$ and similarly, $K(b,t)=0$. More specifically,
consider the normalized $n$-th moment of the Gaussian integral ($%
n=0,1,2$) 
\[ \alpha^nm_n(\alpha,a,b,y)= \left\{ \frac \alpha {i\pi }\right\}^{1/2}
\int_a^b(x-y)^n\exp \left\{ -i\alpha (x-y)^2\right\} dx, \]
where 
\[ \frac m{2\hbar \Delta t}=\alpha.  \]
We change variable to 
\begin{eqnarray}
\sqrt{\alpha }(x-y)=u  \label{subst}
\end{eqnarray}
and get 
\[m_n(\alpha,a,b,y)=\left\{\frac1{i\pi}\right\}^{1/2}\int_{\sqrt{\alpha}
\left(a-y\right)}^{\sqrt{\alpha}\left(b-y\right)}u^n\exp\left\{-iu^2\right\} du.\]
Note that the limits of integration become 
\[ \sqrt{\alpha }(a-y)\to-\infty,\quad \sqrt{\alpha }( b-y)\to\infty\quad%
\mbox{as $\alpha\to\infty$} \]
so that in the limit $\Delta t\rightarrow 0$, that is, as $\alpha \to \infty 
$ and for $a<y<b$, we obtain 
\begin{eqnarray}
m_n(a,b,y)=\lim_{\alpha\to\infty}m_n(\alpha,a,b,y)&=& \int_{\sqrt{\alpha }%
\left( a-y\right) }^{\sqrt{\alpha }\left( b-y\right) }u^n\exp \left\{
-iu^2\right\} du  \nonumber \\
&=&\int_{-\infty }^\infty u^n\exp\left\{ -iu^2\right\}\,du.  \label{moments}
\end{eqnarray}
Note that $m_1=0$ while $m_0=m_2=1$. Now, transferring $K_{N-1}$ to the left
hand side of equation (\ref{Taylor}), dividing by $\Delta t$, taking the
limit, and applying eq.(\ref{moments}), we recover the Schr\"{o}dinger
equation.

At the boundary point $y=b$, the substitution (\ref{subst}) transforms the
upper limit of integration to $0$. In the limit $\Delta t\to 0$, we obtain 
\begin{eqnarray}
m_n(a,b,b)=\lim_{\alpha \to \infty }m_n(\alpha ,a,b,b) &=&\lim_{\alpha \to
\infty }\int_{\sqrt{\alpha }\left( a-b\right) }^0u^n\exp \left\{
-iu^2\right\} \,du  \nonumber \\
&=&\int_{-\infty }^0u^n\exp \left\{ -iu^2\right\} \,du.  \label{momentsb}
\end{eqnarray}
That is, the Fresnel integrals are extended only over half the line. Thus,
we obtain that the coefficient of $K_{N-1}$ in eq.(\ref{Taylor}) is 
\[ m_0(a,b,b)=\frac 12, \]
while that of the first derivative of $K_{N-1}$ is 
\[ \alpha ^{-1}m_1(\alpha ,a,b,b)\to 0 \]
and that of the second derivative of $K_{N-1}$ is 
\[ \alpha ^{-2}m_2(\alpha ,a,b,b)\to 0.  \]
Thus, setting $x=b$ in eq.(\ref{Taylor}) and taking the limit $\Delta t\to 0
$, we obtain eq.(\ref{half}), which in turn implies the boundary condition
eq.(\ref{bc}).

When $a<x<b$, the derivation given in \cite{Keller} leads to the 
Schr\"{o}dinger equation. Thus 
\begin{eqnarray}
{i\hbar }\frac{\partial K(x,t)}{\partial t} &=&-\frac{\hbar ^2}{2m}\,\frac{%
\partial ^2K(x,t)}{\partial x^2}+V(x)K(x,t)\quad \mbox{ for $a<x<b$}
\label{SE} \\
&&\mbox{}  \nonumber \\
K(a,t) &=&K(b,t)=0\quad \quad \mbox{for $t>0$}  \label{bc} \\
&&\mbox{}  \nonumber \\
K(x,0) &=&\delta (x-x_I)\quad \quad \mbox{for $a<x<b$}.  \label{ic}
\end{eqnarray}

In the case $a=-\infty ,\ b=\infty $ every $x$ is an internal point so that
the Schr\"{o}dinger equation is satisfied for all $x$ and thus the Feynman
integral (\ref{FIab}) is equivalent to the Schr\"{o}dinger equation (\ref{SE})
and the initial condition (\ref{ic}) on the entire real line.

This result shows that bounded trajectories imply in effect an infinite
potential barrier on the boundary. In contrast, the Wiener integral over
bounded trajectories leads to the diffusion equation with zero boundary
condition and to the decay of the population. This means that the boundary
is absorbing. This difference between the Wiener and the Feynman integrals
over the same class of functions is illustrative of the different roles that
trajectories play in quantum and classical theories.

The same result has been derived in \cite{Kleinert} and \cite
{Farhi}. In \cite{Kleinert} trajectories are extended into the entire line
as periodic functions. The analysis of this section provides a simpler
derivation of the result. The derivation in \cite{Farhi} is done by
converting the Wiener integral into the Feynman integral by analytic
continuation of $t$ into the imaginary axis.\newline

\noindent
{\large {\bf 4. Feynman integrals with absorbing boundaries}}\\
\renewcommand{\theequation}{4.\arabic{equation}} \setcounter{equation}{0}

Assume now that a trajectory that reaches the boundary $x=a$ or $x=b$ for
the first time is instantaneously absorbed. This means that the wave
function outside the interval $(a,b)$ vanishes identically and the
population inside the interval $(a,b)$ is reduced at a rate determined by
the current at the boundary, as described below. The vanishing wave function
outside the interval $(a,b)$ expresses the assumption that once outside the
interval the particle no longer participates in the quantum evolution of the
particles inside the interval. This may occur, for example, in the
scattering of particles on a target ({\em e.g.}, a nucleus). Particles absorbed in
the nucleus are discounted from the scattered population. Another example is
that of a particle that enters a bath, such as a photographic plate, and
leaves an irreversible trace. Consequently, its quantum interaction with the
particles inside the interval becomes negligible. Also in this case the
population inside the interval is discounted at the rate particles are
absorbed.

The absorption process at a given time $t$ is described as the limiting
process as $\Delta t\rightarrow 0$ of the propagation of a trajectory that
survived in the interval $\left[ a,b\right] $ till time $t$ to a boundary
point in the time interval $\left[ t,t+\Delta t\right] $. In order to
incorporate this behavior into the Feynman formulation, we adopt the
procedure that leads to the Feynman-Kac formula for the probability density
function for a diffusion process with a killing (absorption) measure (see,
{\em e.g.}, \cite{book}). We consider separately the trajectories that reach the
boundary in the time interval $\left[ 0,\Delta t\right] $ , then those that
survived till $\Delta t$, but reach the boundary in the time interval 
$\left[\Delta t,2\Delta t\right]$, and so on. In each time step, the total
population of trajectories has to be discounted by the probability of the
absorbed trajectories. This leads to a modified expression for the
discretized Feynman integral.

First, we calculate the discretized Feynman integral in the time
interval $\left[0,\Delta t\right]$,
\[ \psi_1(x,\Delta t)=\left\{ \frac m{2\pi i\hbar \Delta t}\right\}
^{1/2}\int_a^b\psi _0(x_0)\exp \left\{ \frac i\hbar {\cal S}(x_0,x,\Delta
t)\right\} \,dx_0. \]
Therefore, the probability density of finding a trajectory at $x=a$ in the
time interval $\left[ 0,\Delta t\right] $ is $\left| \psi(a,\Delta
t)\right| ^2$, and there is an analogous expression for the probability
density of finding a trajectory at $x=b$ in the time interval $\left[
0,\Delta t\right] $.

For simplicity, we assume that $b=0$ and consider the interval
$[-a,0]$, where $a>0$. In any time interval $[t,t+\Delta t]$, 
the probability density propagated 
from this interval into the absorbing boundary at $x=0$ is
calculated next. We begin with an initial wave function $\psi (x,t)$ that 
is a polynomial 
\[ Q(x,t)=\sum_{j=1}^Nq_j(t)x^j \]
in the interval $[-a,0]$, such that $Q(-a,t)=Q(0,t)=0$ and $\psi (x,t)=0$
otherwise. The free propagation from the interval $[-a,0]$ is given by 
\[\psi (y,t+\Delta t)=\sqrt{\frac m{2\pi i\hbar \Delta t}}\int_{-a}^0Q(x,t)\exp
\left\{ \frac{im(x-y)^2}{2\hbar \Delta t}\right\} \,dx. \]
The boundary condition at the left end of the support of $\psi (x,t)$ is
written explicitly as 
\[ \sum_{j=1}^Nq_j(t)\left( -a\right) ^j=0. \]
Setting 
\[ \alpha =\frac{\hbar \Delta t}m, \]
the probability mass propagated freely into the absorbing boundary point $x=0$
in time $\Delta t$ is given by 
\[\left| \psi \left( 0,t+\Delta t\right) \right| ^2=\frac
1{2\pi \alpha }\left| \int_{-a}^0Q(x,t)e^{ix^2/2\alpha }\,dx\right| ^2. \]
We change variable by setting $x=\sqrt{\alpha }\xi$ to get 
\begin{equation}
\left| \psi \left( 0,t+\Delta t\right) \right| ^2=\frac{1}{2\pi}
\left| \int_{-a/\sqrt{\alpha }}^0Q\left( \sqrt{\alpha}\xi 
\right) e^{i\xi^2/2}\,d\xi \right|^2 .  \label{integral}
\end{equation}
First, we evaluate the inner integral, 
\[I_N=\sum_{j=1}^Nq_j\sqrt{\alpha^j}\int_{-a/\sqrt{\alpha}}^0
\xi^je^{i\xi ^2/2}\,d\xi . \]
All limits of the type
\[\lim_{\alpha\to0+}\alpha^{-(j+1)/2}
\int_{-a\sqrt{\alpha}} ^0
x^je^{ix^2/2\alpha}\,dx, \]
are understood in the sense
\begin{eqnarray}
\lim_{\alpha\to0+}\alpha^{-(j+1)/2}\int_{-a\sqrt{\alpha}} ^0
x^je^{ix^2/2\alpha}\,dx=\lim_{\epsilon\to0+}\lim_{\alpha\to0+}\alpha^{-(j+1)/2}
\int_{-a\sqrt{\alpha}} ^0 x^je^{(-\epsilon+i)x^2/2\alpha}\,dx. \label{sense}
\end{eqnarray}
The first term in the sum $I_N$ gives
\[ -iq_1(t)\sqrt{\alpha }\int_{-a/\sqrt{\alpha }}^0de^{i\xi^2/2}=
-iq_1(t)\sqrt{\alpha }\left(1-e^{ia^2/2\alpha}\right). \]
The second term gives
\[-iq_2(t)a\sqrt{\alpha}e^{a^2/2\alpha}+iq_2(t)\alpha
\int_{-a/\sqrt{\alpha}}^0e^{i\xi^2/2}\,d\xi.\]
Setting
\[S_j=\alpha^{j/2}\int_{-a/\sqrt{\alpha}}^0\xi^je^{i\xi^2/2}\,d\xi,\]
integration by parts gives  the recursion relation
\[S_j=i\sqrt{\alpha}(-a)^{j-1}e^{ia^2/2\alpha}+i(j-1)\alpha S_{j-2}.\]
Proceeding by induction, we find that for $j>2$
\[S_j=O\left(\sqrt{\alpha}e^{ia^2/2\alpha}\right).\]

Now, using the definition (\ref{sense}), we find that
\begin{eqnarray}
\lim_{\Delta t\to0}\frac{1}{\alpha}
\left|\psi\left(0,t+\Delta t\right)\right|^2=
\frac{1}{2\pi}\left|q_1(t)\right|^2.\label{dpsipx1}
\end{eqnarray}
Keeping in mind that
\[q_1(t)=\frac{\partial\psi(0,t)}{\partial x},\]
we can write
\begin{eqnarray}
\left|\psi\left(0,t+\Delta t\right)\right|^2=\frac{\hbar\Delta t}{2\pi m}
\left|\frac{\partial\psi(0,t)}{\partial x}\right|^2+o\left(\frac{\hbar\Delta
t}{2\pi m}\right)\quad\mbox{for $\alpha\ll1$}.\label{dpsipx}
\end{eqnarray}

Returning to the propagation from an interval $[a,b]$ to its absorbing
boundaries, the Feynman trajectories in the time interval $[0,\Delta t]$ 
consist of those that have not reached the absorbing
boundaries and of those that have. We calculate the Feynman integral
separately on each one of these two classes of trajectories. The integral
over the first class is the one calculated in Section 3 above. That
over the latter class is the integral calculated in the above paragraphs in
this section. According to our assumptions, trajectories that propagate
into the absorbing boundary never return into the interval $[a,b]$ so that the
Feynman integral over these trajectories is supported outside the
interval. On the other hand, the Feynman integral over the bounded
trajectories in the interval is supported inside the interval. Thus 
the two integrals are orthogonal and give rise to no interference.

The Feynman integral over the bounded trajectories represents the wave
function conditioned on not exiting the interval in $\left[0,\Delta
t\right]$. According to eq.(\ref{dpsipx}) and to our assumptions,  
the probability defined by the Feynman integral over the trajectories 
that propagated into the absorbing boundary in this time interval
is given by 
\begin{eqnarray*}
P_1(\Delta t) &=&\lambda (a)\left| \psi_1(a,\Delta t)\right|^2+\lambda
(b)\left|\psi_1(b,\Delta t)\right| ^2 +o(\Delta t)\\
&=&\frac{\hbar \Delta t}{2\pi m}\left\{ \lambda (a)\left| \frac{\partial
\psi_1(a,0)}{\partial x}\right| ^2+\lambda (b)\left| \frac{\partial 
\psi_1(b,0)}{\partial x}\right| ^2\right\} +o\left( \Delta t\right) ,
\end{eqnarray*}
where $\lambda (a)$ and $\lambda (b)$ are the characteristic lengths at $x=a$
and $x=b$, respectively. The survival probability in this time interval is
\begin{eqnarray}
S_1(\Delta t)=1-P_1(\Delta t).\label{S1P1}
\end{eqnarray}

Next, we calculate the discretized Feynman integral in the time interval
$\left[ \Delta t,2\Delta t\right]$ and again break it into the same classes as
above. The discretized integral over the class of bounded trajectories 
in the interval $\left[ a,b\right]$ is
\[\psi_2\left(x,2\Delta t\right) =
\left\{ \frac m{2\pi i\hbar\Delta t}\right\}^{1/2}\int_a^b\psi_1(x_1,\Delta t)
\exp\left\{\frac i\hbar{\cal S}(x_1,x,\Delta t)\right\} \,dx_1. \]
The probability calculated from the Feynman integral over the absorbed
trajectory in the time interval $\left[ \Delta t,2\Delta t\right] $ is 
\[ P_2(2\Delta t)=\lambda (a)\left| \psi_2(a,2\Delta t)\right| ^2+\lambda
(b)\left| \psi_2(b,2\Delta t)\right|^2+o(\Delta t).\]
This is the conditional probability of trajectories that propagate into the absorbing
boundaries in the time interval $\left[ \Delta t,2\Delta t\right]$, given that
they did not reach the boundaries in the previous time interval.
 Thus the discretized survival probability in the time interval $\left[
0,2\Delta t\right]$ is
\begin{eqnarray*}
S_2(2\Delta t)= ( 1-P_1(\Delta t))(1-P_2(2\Delta t))
\end{eqnarray*}

Proceeding this way, we find that the discretized Feynman integral over the class
of bounded trajectories in the time interval $\left[ 0,N\Delta t\right]$ 
is
\begin{eqnarray}
&&\psi_N(x,N\Delta t)=\label{psint}\\ 
&&\left\{ \frac m{2\pi i\hbar\Delta t}\right\}^{1/2}\int_a^b\psi_{N-1}(x_{N-1},
(N-1)\Delta t)\exp\left\{\frac i\hbar{\cal S}(x_{N-1},x,\Delta t)\right\}
\,dx_{N-1}.\nonumber
\end{eqnarray}
As in eq.(\ref{recursion}), we find that $\psi_N(x,N\Delta
t)\to\psi(x,t)$ as $N\to\infty$, where $\psi(x,t)$ is the solution of
Schr\"odinger's equation in $(a,b)$ with the boundary conditions
$\psi(a,t)=\psi(b,t)=0$.

The probability that propagates into the absorbing walls
in the time interval $[(j-1)\Delta t,j\Delta t]$ is given by
\begin{equation}
P_j(j\Delta t)=\lambda (a)\left| \psi _j(a,j\Delta t)\right| ^2\,+\lambda
(b)\left| \psi _j(b,j\Delta t)\right| ^2+o(\Delta t).  \label{pab}
\end{equation}
It follows that the survival probability of trajectories inside the
interval is
\begin{equation}
S(t)=\lim_{N\rightarrow \infty }\prod_{j=1}^N\left( 1-P_j(j\Delta
t)\right) .  \label{sp}
\end{equation}
According to eqs.(\ref{dpsipx}) and (\ref{pab}), $P_j(j\Delta t)$ is given by 
\[P_j(j\Delta t)=\frac{\hbar \Delta t}{2\pi m}\left[ \lambda (a)\left| \frac
\partial {\partial x}\psi_{j-1}\left( a,(j-1)t\right) \right| ^2+\lambda
(b)\left| \frac \partial {\partial x}\psi_{j-1}\left( b,(j-1)t\right) \right|
^2+o(1)\right] , \]
so that eq.(\ref{sp}) gives  the survival probability 
\begin{eqnarray}
S(t)=  
\exp \left\{ -\frac \hbar {\pi m}\int_0^t\left[ \lambda (a)\left| \frac
\partial {\partial x}\psi\left( a,t^{\prime }\right) \right| ^2\,+\lambda
(b)\left| \frac \partial {\partial x}\psi\left( b,t'\right) \right|
^2\right] \,dt'\right\} . \label{1-pt} 
\end{eqnarray}

The wave function of the trajectories that have not been absorbed by time $t$
is the wave function of a particle, conditioned on not reaching the
absorbing boundary by time $t$. The conditioning renormalizes the wave
function inside the domain at all times and thus it remains $\psi\left(
x,t\right) $. The survival probability is $1-P(t).$ If a particle is known
not to have been absorbed by time $t_1$, its survival probability till time $%
t_2>t_1$, denoted $S(t_2,t_1)$, is given by 
\begin{equation}
S(t_2,t_1)=\exp \left\{ -\frac \hbar {\pi m}\int_{t_1}^{t_2}\left[ \lambda
(a)\left| \frac \partial {\partial x}\psi\left( a,t\right) \right| ^2\,+\lambda
(b)\left| \frac \partial {\partial x}\psi\left( b,t\right) \right| ^2\right]
\,dt\right\} .  \label{S12}
\end{equation}

The survival probability $S(t)$ and the wave function $\psi(x,t)$ can be
combined into a discounted wave function
\begin{eqnarray}
\Psi(x,t)=\sqrt{S(t)}\,\psi(x,t).\label{PSI1}
\end{eqnarray}
The discounted wave function should be used for all purposes if an
absorbing surface is present in the system. It is normalized to $S(t)$ and
decays in time. The the wave function $\psi(x,t)$ is recovered from
$\Psi(x,t)$ by conditioning on survival of the particle by time $t$.\\

\noindent
{\large {\bf 5. Other modes of absorption}}\newline
\renewcommand{\theequation}{5.\arabic{equation}} \setcounter{equation}{0}

There are different phenomenological descriptions of absorption and
decay in the literature. These may include different absorption rates at
different energies, absorption of two non-interacting
particles at a wall, and so on. It is not {\em a priori} obvious which
mode of absorption applies in a given physical situation without a
detailed analysis of the absorption mechanism. We show below that some of
the phenomenological descriptions of absorption can be derived from our
formalism.\newline

\noindent{\bf 5.1 Absorption in energy windows}\newline

Some of the descriptions are based on the premise that each mode is absorbed
(or decays) at a rate proportional to its energy. This type of absorption
can be obtained from the model of absorption of Feynman trajectories at an
absorbing wall. Since the trajectories that reach the wall in time $\Delta t$
cannot be separated into those that belong to one mode or another, the
overall discounted probability has to be a function of the rate and
probability of propagation of trajectories into the wall and the overall
energy of the system. The way the discounting is spread among the different
components of the surviving wave function can be chosen according to any
number of criteria.

Therefore, we adopt the following absorption principles: the overall
probability discounted at each time step is a function of the rate of
propagation of probability into the wall and the average energy of the
system; the probability discounted from each mode is, in addition, a
function of the energy of that mode. According to these principles, the
factor $\lambda $ is allowed to be a function of the propagation rate, that
is, 
\[ \lambda =\lambda \left( \frac \hbar {\pi m}\left| \frac \partial {\partial
x}\psi\left( 0,t\right) \right| ^2\right) . \]
In addition, it can depend on
the average energy, $\left\langle {\cal E}\right\rangle $, defined
as follows. Consider a single absorbing wall placed at the origin and assume
that the wave function in the presence of an infinite potential wall at the
origin is
 \[\psi\left(x,t\right)=\sum_{k=1}^\infty a_k\psi_k\left(x\right)\exp
\left\{-\frac i\hbar E_kt\right\},\]
where the eigenfunctions $\psi _k\left( x\right) $, corresponding
to energies $E_k$, form a complete set. The sum is replaced with an 
integral if the spectrum is continuous. The normalization condition is 
\[ \sum_{k=1}^\infty \left| a_k\right| ^2=1. \]
The energy of a mode is defined as 
\begin{eqnarray}
{\cal E}_k=\left| a_k\right| ^2E_k\label{Ek}
\end{eqnarray}
and we define the average energy per mode as 
\[\left\langle{\cal E}\right\rangle=\sum_{k=1}^\infty\left|a_k\right|^2
{\cal E}_k=\sum_{k=1}^\infty\left|a_k\right|^4E_k.\]
Thus the factor $\lambda $ has the form 
\[
\lambda =\lambda \left( \frac \hbar {\pi m}\left| \frac \partial {\partial
x}\psi\left( 0,t\right) \right| ^2,\left\langle {\cal E}\right\rangle \right) . 
\]
The overall probability propagating into the wall in the time interval $%
\left[ t,t+\Delta t\right] $ is given by 
\begin{equation}
P\left( t\right) =\frac{\lambda \hbar }{\pi m}\left| \frac \partial
{\partial x}\psi\left( 0,t\right) \right| ^2\,\,\Delta t+O\left( \Delta
t^{3/2}\right) .  \label{overall}
\end{equation}

Now, we spread this discounted probability among the different modes. The
mode $k$ is discounted by 
\[
a_k\psi _k\left( x\right) \exp \left\{ -\frac i\hbar E_kt\right\} \left( 1-%
\frac{\lambda _k}2\frac \hbar {\pi m}\left| \frac \partial {\partial
x}\psi\left( 0,t\right) \right| ^2\,\,\Delta t+O\left( \Delta t^{3/2}\right)
\right) , 
\]
where $\lambda _k$ is defined below. The value of $\lambda _k$ has to be
chosen so that the overall discounted probability is as given in eq.(\ref
{overall}). The total surviving population is 
\begin{eqnarray*}
&&\int_{-\infty }^0\left| \sum_{k=1}^\infty a_k\psi _k\left( x\right) \exp
\left\{ -\frac i\hbar E_kt\right\} \left( 1-\frac{\lambda _k}2\frac \hbar
{\pi m}\left| \frac \partial {\partial x}\psi\left( 0,t\right) \right|
^2\,\,\Delta t+O\left( \Delta t^{3/2}\right) \right) \right| ^2\,dx \\
&=&\sum_{k=1}^\infty \left| a_k\right| ^2\left( 1-\lambda _k\frac \hbar {\pi
m}\left| \frac \partial {\partial x}\psi\left( 0,t\right) \right| ^2\,\,\Delta
t+O\left( \Delta t^{3/2}\right) \right) ,
\end{eqnarray*}
so that the overall discounted probability is 
\[
\sum_{k=1}^\infty \left| a_k\right| ^2\lambda _k\frac \hbar {\pi m}\left|
\frac \partial {\partial x}\psi\left( 0,t\right) \right| ^2\,\,\Delta t+O\left(
\Delta t^{3/2}\right) =\frac{\lambda \hbar }{\pi m}\left| \frac \partial
{\partial x}\psi\left( 0,t\right) \right| ^2\,\,\Delta t+O\left( \Delta
t^{3/2}\right) . 
\]
Simplification leads to 
\begin{equation}
\sum_{k=1}^\infty \left| a_k\right| ^2\lambda _k=\lambda .
\label{lnormalization}
\end{equation}
We set 
\[
\lambda _k=\lambda \alpha _k 
\]
so that eq.(\ref{lnormalization}) becomes 
\[
\sum_{k=1}^\infty \left| a_k\right| ^2\alpha _k=1. 
\]
To obtain absorption rates proportional to energies, we have to choose 
\[
\alpha _k=\frac{{\cal E}_k}{\left\langle {\cal E}\right\rangle }. 
\]
It follows that 
\begin{equation}
\lambda _k=\lambda \frac{{\cal E}_k}{\left\langle {\cal E}\right\rangle }
\label{lke}
\end{equation}
and the absorption law for each mode reduces to 
\[
\exp \left\{ -\int_0^t\lambda \frac{{\cal E}_k}{\left\langle {\cal E}%
\right\rangle }\frac \hbar {\pi m}\left| \frac \partial {\partial x}\psi\left(
0,t^{\prime }\right) \right| ^2\,dt^{\prime }\right\} . 
\]
Now, we choose $\lambda $ so that 
\[ \lambda \frac \hbar {\pi m}\left| \frac \partial {\partial x}\psi\left(
0,t\right) \right| ^2=\frac{\left\langle {\cal E}\right\rangle }\hbar , \]
we obtain the discounted wave function of the surviving trajectories in the
form 
\begin{eqnarray}
\Psi \left( x,t\right) =\sum_{k=1}^\infty a_k\psi _k\left( x\right) \exp
\left\{ -\frac i\hbar E_kt\right\} \exp \left\{ -\frac{{\cal E}_k}{2\hbar}
t\right\}. \label{PSIE}
\end{eqnarray}
The survival probability of mode $k$ is
$$ S_k(t)=|a_k|^2\,\exp \left\{ -\frac{{\cal E}_k}{\hbar} t\right\}$$
and the total survival probability is the sum
\begin{eqnarray}
S_{{\cal E}}(t)=\int_{-\infty}^\infty\left|\Psi(x,t)\right|^2\,dx=
\sum_{k=1}^\infty |a_k|^2\,\exp \left\{ -\frac{{\cal E}_k}{\hbar}
t\right\}.\label{S_E}
\end{eqnarray}
This mode of decay is similar to that in diffusion, where each mode
decays separately at a different decay rate. 

Now, we consider an interface between two media, one of which absorbs
particles with energies in a given band $E_L<E<E_U$ and reflects all
particles with energies outside the band. This situation is described by an
infinite potential at the interface for particles with energies outside the
band and an absorbing wall for particles with energies inside the band.

According to the formalism developed above, all particles see effectively an
infinite wall at the interface, but only those with energies in the band are
absorbed. To describe the absorption process, we separate the initial wave
function $\psi _0(x)$ into its projection, $\psi _{band}(x)$, on the
subspace spanned by eigenfunctions corresponding to the infinite wall whose
eigenvalues (energies) are in the band, and to its projection, $\psi _c(x)$,
on the orthogonal complement of this subspace. The wave function, $\psi
(x,t) $, is also decomposed into analogous projections, 
\[ \psi (x,t)=\psi _{band}(x,t)+\psi _c(x,t), \]
where $\psi _{band}(x,t)$ decays according to the above absorption
formalism, whereas $\psi _c(x,t)$ evolves according to the Schr\"{o}dinger
equation with zero boundary conditions on the interface. The $\lambda _k$
for this case are chosen by eq.(\ref{lke}) for energies in the band and $0$
outside the band. The mean energy per mode is defined only for the modes
that are absorbed, that is, 
\[\left\langle{\cal E}\right\rangle
=\sum_{k\in\mbox{band}}\left|a_k^4E_k\right|.\]

\vspace{3mm}

\noindent {\bf 5.2 Absorption of two non-interacting particles at a wall}%
\newline

The Hamiltonian for two non-interacting particles has the structure 
\[H\left( x_1,p_1;x_2,p_2\right) =H_1\left( x_1,p_1\right) +H_2
\left(x_2,p_2\right) \]
so that the wave function has the form
\[ \psi \left( x_1,x_2,t\right) =\psi _1(x_1,t)\psi _2(x_2,t). \]
If an absorbing wall for both particles is placed at $x=0$ and
initially both
particles are to the left of the wall, the configuration space is the
third quadrant in the $\left( x_1,x_2\right) $-plane. The absorbing
boundary, denoted $\partial D$, consists of the negative $x_1$ and $x_2$
axes. Each particle can have a different $\lambda$ so that a 
two dimensional analog of the calculation of the absorption rate
gives the instantaneous rate 
\begin{eqnarray}
&&\frac{\lambda_1\hbar}{\pi m_1}
\int_{-\infty}^0\left|\frac{\partial\psi_1(0,t)}{\partial x_1}
\right|^2\left|\psi_2(x_2,t)\right|^2\,dx_2+
\frac{\lambda_2\hbar}{\pi m_2}\int_{-\infty }^0\left| 
\frac{\partial\psi_2(0,t)}{\partial x_2}\right|^2\left|\psi_1(x_1,t)
\right|^2\,dx_1=\nonumber\\
&&\mbox{}\label{2indep}\\
&&\frac{\lambda_1\hbar}{\pi m_1}\left|\frac{\partial\psi_1(0,t)}
{\partial x_1}\right|^2+
\frac{\lambda_2\hbar}{\pi m_2}\left|\frac{\partial\psi_2(0,t)}
{\partial x_2}\right|^2. \nonumber
\end{eqnarray}
Thus the discounted wave function is given by
\begin{eqnarray*}
\Psi\left( x_1,x_2,t\right) =\Psi_1(x_1,t)\Psi_2(x_2,t),
\end{eqnarray*}
where
\begin{eqnarray*}
\Psi_j(x_j,t)=\psi_j(x_j,t)\exp\left\{\frac{\lambda_j\hbar}{\pi
m_j}\left|\frac{\partial\psi_j(0,t)}
{\partial x_j}\right|^2\right\},\quad(j=1,2).
\end{eqnarray*}
Thus, the population decays together at a rate that is the sum of the
two rates. The decay law is different than that of a single particle
in a superposition of two states in that there are no beats in the
decay law of two independent particles.

It is apparent from the above examples that different modes of decay
can be obtained from the model of an absorbing wall. It is not
a-priory clear which mode of decay is appropriate for a given physical
system. The choice of the decay mode depends on the particular
physical system.\\

\noindent
{\large{\bf 6. The absorption current and the slit experiment}}\newline
\renewcommand{\theequation}{6.\arabic{equation}} \setcounter{equation}{0}

An absorbing boundary engenders an absorption probability current
 into the boundary \cite{MSU}. The simplest definition of the absorption
current at a single absorbing point, $x=0$, in one dimension, is (see
eq.(\ref{1-pt})) 
\begin{eqnarray}
{\cal J}(0,t) &=&\frac d{dt}\left[ 1-S(t)\right] =  \label{udc} \\
\ \, &=&\frac{\lambda\hbar}{\pi m}\left| \frac{\partial \psi(0,\ t)}{\partial x}\right|
^2\exp \left\{ -\frac{\lambda\hbar}{\pi m} \int_0^t\ \left| \frac{\partial \psi(0,t^{\prime
})}{\partial x}\right| ^2\,dt^{\prime }\right\} ,  \nonumber
\end{eqnarray}

In higher dimensions the discounted wave function in a domain ${\cal D}$ in 
the presence of an absorbing boundary $\Gamma $ is given by 
\begin{equation}
\Psi ({\bf x},t)=\psi ({\bf x},t)\exp \left\{ -\frac{\lambda\hbar}{2\pi m}
\int_0^t\!\oint_\Gamma \left| \frac{\partial \psi ({\bf x}^{\prime },t^{\prime })}{%
\partial {\bf n}}\right| ^2\,\,dS_{{\bf x}^{\prime }}\,dt^{\prime }\right\} ,
\label{PSI}
\end{equation}
where $\psi ({\bf x},t)$ is the solution of Schr\"{o}dinger's
equation in ${\cal D}$ with zero boundary condition on $\Gamma$ and ${\bf n}$
is the unit outer normal to the boundary. Adopting
the interpretation of the squared modulus of the wave function as the
probability density of finding a particle at the point ${\bf x}$ at time $t$
(whatever that means), the discounted wave function can be used to calculate
the {\em joint} probability density of surviving by time $t\,$and finding
the particle at ${\bf x}$ at the same time. The squared modulus of the wave
function, {\em conditioned} on surviving by time $t$ is found by dividing
the joint probability density $\left| \Psi ({\bf x},t)\right| ^2$ by the
probability of the condition, $S(t)$. In the multi-dimensional case at hand 
\begin{equation}
S(t)=\exp \left\{ -\frac{\lambda\hbar}{\pi m} \int_0^t\ \oint_\Gamma \left| \frac{\partial \psi
({\bf x}^{\prime },t^{\prime })}{\partial {\bf n}}\right| ^2\,\,dS_{{\bf x}%
^{\prime }}\,dt^{\prime }\right\} .  \label{mdSt}
\end{equation}
It follows from eqs.(\ref{PSI}) and (\ref{mdSt}) that the conditioned wave
function is $\psi ({\bf x},t)$.

Generalizing the definition (\ref{udc}) to higher dimensions, the total
current at the absorbing boundary is defined as
\begin{eqnarray}
{\cal J}_\Gamma(t) &=&\frac d{dt}\left[ 1-S(t)\right]\label{mudc} \\
&=&\frac{\lambda\hbar}{\pi m} \oint_\Gamma\left|\frac{\partial\psi({\bf x},t)}{\partial{\bf n}}
\right|_\Gamma^2\,dS_{{\bf x}'}\,\exp\left\{-\frac{\lambda\hbar}{\pi m} 
\int_0^t\oint_\Gamma\left| 
\frac{\partial\psi({\bf x}',t')}{\partial {\bf n}}\right|^2\,dS_{{\bf x}'}\,dt'
\right\}.  \nonumber
\end{eqnarray}

The normal component of the multi-dimensional probability current density 
at any point ${\bf x}$ on $\Gamma $ is given by 
\begin{equation}
\left. {\cal J}({\bf x},t)\cdot {\bf n}\left( {\bf x}\right) \right| _\Gamma
=\frac{\lambda\hbar}{\pi m} 
\left| \frac{\partial \psi ({\bf x},t)}{\partial {\bf n}}\right|
_\Gamma ^2\,\exp \left\{ -\frac{\lambda\hbar}{\pi m} 
\int_0^t\oint_\Gamma \left| \frac{\partial
\psi ({\bf x}^{\prime },t^{\prime })}{\partial {\bf n}}\right| ^2\,\,dS_{%
{\bf x}'}\,\,dt'\right\} .  \label{JnG}
\end{equation}

Next, we consider the slit experiment with an absorbing screen ({\em e.g.},
photographic plate or a fluorescent screen). In this case, unlike the usual
assumption in quantum mechanics \cite{Feynman}, the pattern that appears on
the absorbing screen cannot be the squared modulus of the wave function in the
entire space (with or without a finite potential), as the above analysis
implies. It cannot be the modulus of the wave function with an absorbing
boundary because the wave function vanishes on an absorbing boundary. According
to the absorption principles, the pattern on the screen represents the
probability density that propagates into the screen. This probability density
is the unidirectional absorption current on the screen. Thus, the instantaneous
pattern at time $t$ is $\left.{\cal J}({\bf x},t)\cdot{\bf n}\left({\bf x}
\right)\right|_\Gamma$ and the cumulative pattern, after the arrival of many
particles, is
\begin{eqnarray}
\left.{\cal J}({\bf x})\cdot{\bf n}\left({\bf x}\right)\right|_\Gamma
=\int_0^\infty\left.{\cal J}({\bf x},t)\cdot {\bf n}
\left({\bf x}\right)\right|_\Gamma\,dt.\label{Jx}
\end{eqnarray}
The instantaneous pattern $\left.{\cal J}({\bf x},t)\cdot{\bf n}\left({\bf x}
\right)\right|_\Gamma$ is obtained as the histogram of points on the screen
collected at time $t$ after releasing the particle in the slit. The cumulative
current density $\left.{\cal J}({\bf x})\cdot{\bf n}\left({\bf x}\right)
\right|_\Gamma$ is the histogram of points on the screen collected at all times.

We consider the following experimental setup for the slit experiment with an
absorbing screen. A planar screen is placed in
the plane $x=0$ and another screen it placed in the plane $x=x_0$ and it is
slit along a line parallel to the $z-$axis. Due to the invariance of the
geometry of the problem in $z$ the mathematical description of the slit is,
following \cite{Feynman}, an initial truncated Gaussian wave packet in the $%
(x,y)-$plane, concentrated around the initial point, $(x_0,0).$ To describe
the interference pattern on the screen, we assume it is an absorbing line on
the $y$-axis and apply the formalism developed above. The wave function, as
given by our formalism, evolves from the initial packet according to eq.(\ref
{PSI}), as 
\[\Psi (x,y,t)=\psi (x,y,t)\exp \left\{ -\frac{\lambda\hbar}{2\pi m}
\int_0^t\!\int_{-\infty }^\infty \left| \frac{\partial \psi (0,y^{\prime },t^{\prime
})}{\partial x}\right| ^2\,\,dy^{\prime }\,dt^{\prime }\right\} \,, 
\]
where $\psi (x,y,t)$ is the solution of the Schr\"{o}dinger equation in
the half plane $x>0$ with $\psi (0,y,t)=0$ and $\psi (x,y,0)$ is the
given initial packet. 

The probability current density on the screen is given by
\begin{eqnarray}
{\cal J}(0,y,t)=\frac{\lambda\hbar}{\pi m} 
\left| \frac{\partial \psi (0,y,t)}{\partial x}\right|
^2\exp \left\{ -\frac{\lambda\hbar}{\pi m} \int_0^t\!
\int_{-\infty }^\infty \left| \frac{%
\partial \psi (0,y^{\prime },t^{\prime })}{\partial x}\right|
^2\,\,dy^{\prime }\,dt^{\prime }\right\}  \label{Jyt}
\end{eqnarray}
This is the probability density of a collapse of the wave function occurring
at the point $y$ on the screen at time $t$. The total current 
\begin{equation}
{\cal J}(y)=\frac{\lambda\hbar}{\pi m} \int_0^\infty
\left| \frac{\partial \psi (0,y,t)}{\partial x}\right|
^2\exp \left\{ -\frac{\lambda\hbar}{\pi m} \int_0^t\!
\int_{-\infty }^\infty \left| \frac{%
\partial \psi (0,y^{\prime },t^{\prime })}{\partial x}\right|
^2\,\,dy^{\prime }\,dt^{\prime }\right\}\,dt  \label{Jy}
\end{equation}
is the probability density that the collapse of the wave function occurs at
the point $y$ on the screen (ever). If the initial distribution of
velocities in the $x$ direction is concentrated about $v_0$, the density $%
{\cal J}(y)$ is approximately the same as ${\cal J}(0,y,\bar{t})$, where 
\cite{Feynman} 
\[
\bar{t}=\frac{x_0}{v_0}. 
\]

In a real experiment the measurement is neither instantaneous nor infinite
in time. That is, an integral over a finite time interval is observed rather
than (\ref{Jyt}) or (\ref{Jy}). If a packet of particles is sent out eq.(%
\ref{Jyt}) is the probability density of Feynman trajectories that propagate
instantaneously at time $\,t$ into the point $(0,y)$ in the screen and is
seen as the density of light intensity on the (ideal) fluorescent screen at
time $t$. The function (\ref{Jy}) represents the cumulative (in time)
probability current density of Feynman trajectories absorbed in the wall and
is seen as the density of the trace the initial packet eventually leaves on
the screen ({\em e.g.}, photographic plate).

\medskip To determine the patterns (\ref{Jyt}) and (\ref{Jy}), we have to
calculate first the two-dimensional wave function with zero boundary
condition on the $y$-axis. It can be written as 
\[
\psi (x,y,t)=\psi ^1(x,t)\psi ^2(y,t), 
\]
where $\psi ^1(x,t)$ is the function $K(x,t)$ in
eqs.(\ref{SE}, \ref{bc}, \ref{ic}) gaven above and by 
eq.(\ref{wave}) geven below.

Note that according to our theory $\psi^1(0,t)=0$, because the absorbing
 screen is placed at $x=0$. The component $\psi ^2(y,t)$ of the wave function
is given by the freely propagating Gaussian slit \cite{Feynman} 
\[
\psi ^2(y,t)=\int_{-\infty }^\infty \frac{e^{-z^2/2\sigma _y^2}}{\sqrt{2\pi i%
}\sigma _y}\frac{e^{-im(y-z)^2/2\hbar t}}{\sqrt{2\pi i\hbar t/m}}\,dz. 
\]
Evaluation of the integral gives 
\[
\left| \psi ^2(y,t)\right| ^2=\frac 1{2\pi \sigma _y}\left( \displaystyle{%
\frac{\hbar ^2t^2}{\sigma _y^2m^2}+\sigma _y^2}\right) ^{-1/2}\exp \left\{ {%
\displaystyle{-\frac{y^2}{\displaystyle{\frac{\hbar ^2t^2}{\sigma _y^2m^2}%
+\sigma _y^2}}}}\right\} 
\]
If $\sigma _x<<|x_0|$, the upper limit of integration in eq.(\ref{1-pt}) can
be replaced by $\infty $ with a transcendentally small error.

According to eq.(\ref{Jyt}), the instantaneous absorption rate at time $t$
at a point $(0,y)$ on the screen is given by 
\begin{equation}
{\cal J}(0,y,t)=\left| \frac{\partial K(0,t)}{\partial x}\right| ^2\left|
\psi ^2(y,t)\right| ^2,  \label{absrate}
\end{equation}
where $\left| \partial K(0,t)/\partial x\right| ^2$ is given in eq.(\ref
{dK0t2}) below. Thus the pattern on the screen at time $t$ is given by $%
{\cal J}(0,y,t)$ in eq.(\ref{absrate}).

To compare eq.(\ref{absrate}) with that given in \cite{Feynman}, we
reproduce the derivation of \cite{Feynman} with an initial two-dimensional
Gaussian wave packet. The result gives the wave function as 
\[
\psi _F(x,y,t)=\psi _F^1(x,t)\psi ^2(y,t) 
\]
and probability density at the screen at time $t$ as 
\begin{equation}
\left| \psi _F(0,y,t)\right| ^2=\frac 1{2\pi \sigma _x}\left( \frac{\hbar
^2t^2}{\sigma _x^2m^2}+\sigma _x^2\right) ^{-1/2}\exp \left\{ \displaystyle{-%
\frac{x_0^2}{\displaystyle{\ \frac{\hbar ^2t^2}{\sigma _x^2m^2}+\sigma _x^2}}%
}\right\} \left| \psi ^2(y,t)\right| ^2.  \label{psiF0}
\end{equation}
Comparing equation (\ref{psiF0}) with eq.(\ref{absrate}), we see that the
pattern predicted by Feynman's theory differs from that in the
present theory by a time dependent factor only. The instantaneous intensity
of the diffraction pattern in the absence of an absorbing screen, given in 
\cite{Feynman}, is defined as $\left| \psi _F(0,y,t)\right| ^2$. Thus, the
introduction of an absorbing screen, according to these interpretations,
gives the relative brightness as 
\[
\displaystyle{\frac{\displaystyle\lambda \displaystyle
{\left| \displaystyle{\frac{\partial \psi (0,y,t)}{\partial x}}\right| ^2}}{%
\left| \psi _F(0,y,t)\right| ^2}}, 
\]
which is a function of time only. The decay in time of the quotient reflects
the fact that the absorbing screen depresses the entire wave function in
time. Thus, the part of the packet that arrives later is already attenuated
by the preceding absorption, relative to the unattenuated wave function in
the absence of absorption.\newline

\noindent
{\large{\bf 7. Examples}}\newline
\renewcommand{\theequation}{7.\arabic{equation}} \setcounter{equation}{0}

In the following examples possible experiments are discussed, that
can be used to verify the predictions of the above theory in the different modes of
absorption. The examples include a particle between
absorbing walls, a free particle incident on an absorbing wall, and the slit
experiment bounded by absorbing walls. The latter indicates that the pattern
on the screen changes when absorbing walls are present. \\

\noindent
{\bf 7.1 A particle between two absorbing walls}\newline

First, we consider a particle with symmetric absorbing walls at $x=0,a$ and
zero potential. We assume that $\lambda _{-a}=\lambda _a$. The wave function
is given by 
\[\psi(x,t)=\sum_{n=1}^\infty A_n\exp \left\{ -\frac{i\hbar n^2\pi ^2}{2ma^2}%
t\right\} \sin \frac{n\pi }ax.\]
It was shown in \cite{PLA} that
for a particle with a single energy level the wave function decays at an
exponential rate proportional to the energy. However, if there are more than
just one level, the exponent contains beats. For example, for a two level
system with real coefficients, we obtain the survival probability 
\begin{eqnarray}
&&S(t)=\label{survp}\\
&&\exp \left\{ -\frac{\lambda _a\hbar }{m\pi }\left[ \frac{\pi ^2}{a^2}%
\left( A_k^2k^2+A_n^2n^2\right) t-\frac{4m(-1)^{k+n}knA_kA_n}{\hbar \left(
n^2-k^2\right) }\sin \frac{\hbar (n^2-k^2)\pi ^2}{2ma^2}t\right] \right\}.\nonumber
\end{eqnarray}
The strongest beats occur for $k=2,\,n=1$ with frequency $\omega
_{1,2}=3\hbar \pi ^2/2ma^2$. Setting $A_1=A_2=\sqrt{1/2a}$ and introducing
the dimensionless time $\tau =\omega _{1,2}t$, we find that the survival
probability is 
\[
S(t)=\exp \left\{ -\frac{\lambda _a}{3\pi }\left( 5\tau -2\sin \tau \right)
\right\} . 
\]
The function 
\[
\log S(t)=-\frac{\lambda _a}{3\pi }\left( 5\tau -2\sin \tau \right) 
\]
is qualitatively similar to that given in \cite{Wilkinson}.

If absorption in windows of energy is assumed, the
discounted wave function (\ref{PSIE}) is 
\begin{equation}
\Psi (x,t)=\sum_{n=1}^\infty A_n\exp \left\{ -\frac{i\hbar n^2\pi ^2}{2ma^2}%
t\right\} \exp \left\{ -\frac{\left| A_n\right| ^2\hbar n^2\pi ^2}{2ma^2}%
t\right\} \sin \frac{n\pi }ax. \label{kdecay}
\end{equation}
Writing
\[A_n=\frac{a_n}{\sqrt{a}},\]
such that $\sum|a_n|^2=1$, we obtain that
the survival probability (\ref{S_E}) is 
\begin{eqnarray}
 S_E(t)=\sum_{n=1}^\infty \left| a_n\right| ^2\exp \left\{ -\frac{\left|
a_n\right| ^2\hbar n^2\pi ^2}{ma^2}t\right\}. \label{SLSE}
\end{eqnarray}
In the particular case that there is only one energy level in the box,
both results (\ref{survp}) and (\ref{SLSE}) should be the same. In this
case, we obtain
\[\lambda=\pi a.\]

\vspace{3mm}
\noindent 
{\bf 7.2 A wave packet incident on an absorbing wall}\newline

Next, we consider a Gaussian-like wave packet of free particles traveling
toward an absorbing wall at $x=0$ with positive mean velocity $k_0$. The
wave function is given by 
\begin{eqnarray}
&&\Psi \left( x,t\right)=\left(\frac{2a^2}\pi\right)^{1/4}\frac{A^{-1/2}
e^{i\varphi}}
{\left(a^4+\frac{4\hbar^2t^2}{m^2}\right)^{1/4}} \times\label{wave}\\
&&\left[\exp\left\{-\frac{\left(x+x_0-\frac{\hbar k_0}mt\right)
^2}{a^2+\frac{2i\hbar t}m}+ik_0\left( x+x_0\right) \right\}
 -\exp \left\{ -\frac{\left( x-x_0+\frac{\hbar k_0}mt\right)
^2}{a^2+\frac{2i\hbar t}m}-ik_0\left( x-x_0\right) \right\} \right] ,\nonumber
\end{eqnarray}
where
\[A=\left[ 1-\exp \left( -\frac{\ 2x_0^2}{a^2}\right) \exp \left( -\frac
12a^2k_0^2\right) \ \right]\]
and $\varphi$ is a phase (see \cite[p.64]{CT}).
It follows that
\begin{eqnarray}\
&&\frac{\lambda \hbar }{m\pi }\left|\frac{\partial}{\partial x}
 \psi\left( 0,t\right) \right|^2=\label{dK0t2}\\
&&\frac{\sqrt{2}\lambda \hbar am^2\left( 4x_0^2+a^4k_0^2\right) }{\left(
\sqrt{\pi }\right) ^3A\left( \sqrt{a^4m^2+4t^2\hbar ^2}\right) ^3}\exp
\left( -2a^2\frac{x_0^2m^2+\hbar ^2k_0^2t^2-2k_0mx_0t\hbar
}{a^4m^2+4t^2\hbar ^2}\right) ,\nonumber    
\end{eqnarray}
hence 
\begin{equation}
\int_0^\infty\left|\frac\partial{\partial x}\psi(0,t)\right|^2\,dt
<\infty.\label{FINT}
\end{equation}
Thus 
\[ R=S(\infty)=\lim_{t\rightarrow \infty }\left[ 1-P(t)\right] >0, \]
that is, the wave packet is only partially absorbed. This means that the
``reflected'' wave consists of trajectories that turned around
before propagating past the absorbing wall. The absorption occurs when the
trajectory propagates into the medium inside the wall. The discount of the
wave function occurs when the packet is at the wall, as can be seen from
eqs. (\ref{dK0t2}) and (\ref{FINT}). Thus $R$ plays the role of a {\em %
reflection coefficient}. This is neither the usual reflection coefficient
for a finite potential barrier nor that for an infinite barrier. The
reflection coefficient is a function of the packet group velocity, $k_0$,
the width of the packet, $a$, its initial distance from the absorbing
screen $x_0$, the parameter $\lambda$, and its mass $m$. 
This dependence is experimentally measurable.

If absorption in energy windows is assumed, the discounted wave function,
as given in eq.(\ref{PSIE}), is  calculated as follows. The modes are the
functions
$$\psi_k(x)=\sqrt{\frac2\pi}\sin kx$$
and the energies $E_k$ are
$$E_k=\frac{\hbar^2k^2}{2m}.$$
The discounted wave function is given by eq.(\ref{PSIE}) with summation 
replaced by integration with respect to $k$ over the entire real line. 
The coefficient $a_k$ is the sine transform of the initial wave function 
(\ref{wave}) with $t=0$,
\begin{eqnarray*}
&&a_k=-i(2\pi)^{1/4}\sqrt{\frac aA}\times\\
&&\exp \left\{-\frac{x_0^2}{a^2}\right\}\left(\exp\left\{\frac{\left(x_0-
\frac{ika^2+ik_0a^2}2\right)^2}{a^2}\right\}-\exp \left\{
\frac{\left( x_0-\frac{-ika^2+ik_0a^2}2\right)^2}{a^2}\right\} \right)
\end{eqnarray*}
and the decay rate of the $k$-th mode, as given in eqs.(\ref{Ek}),
(\ref{PSIE}), is
\begin{eqnarray*}
\frac{{\cal E}_k}{\hbar}&=&\frac{\left|a_k\right|^2\hbar^2k^2}{2m\hbar}\\
&=&\frac{a\hbar k^2\sqrt{2\pi}}{4mA}
\exp \left\{ -\frac{k_0^2a^2}2\right\} \exp \left\{
-\frac{k^2a^2}2\right\} \left[ \cosh \left( kk_0a^2\right) -\cos \left(
2x_0k\right) \right].
\end{eqnarray*}
According to Section 5.1, the survival probability is the 
same as in the previous case.\\

\noindent 
{\bf 7.3 The slit experiment with lateral absorbing boundaries }\newline

Here, we propose a simple device for performing a measurement of times of
arrival of particles at an absorbing boundary in one dimension, as well as
that of the unidirectional current at the absorbing boundaries. In addition,
the proposed device demonstrates the effect of additional absorbing
boundaries on the slit experiment. The proposed measurement can discriminate
between the various modes of absorption described above.

Consider the setup of the slit experiment enclosed between two parallel
absorbing walls, symmetric with respect to the slit and perpendicular to the
planes of the screen and the slit. For example, the walls can be made of
photographic plates. Particles are given a constant initial velocity, $v_x$,
in the $x$ direction (perpendicular to the planes of the screen and slit),
within the constraints of uncertainty. The time a particle leaves the slit
is also measurable within the constraints of uncertainty.

The initial packet is Gaussian in the $x$ direction and is uniform inside
the slit (in the $y$ direction). This means that the initial velocities in
the $y$ direction have the density 
\begin{equation}
\left| \hat{\Psi}\left( k\right) \right| ^2=\left| \frac{\sin \frac \pi 2k}{%
\frac \pi 2k}\right| ^2.  \label{velocityd}
\end{equation}
The plane of the slit is $x=x_0$, the plane of the screen is $x=0$, the slit
is the interval $-\pi /2<y<\pi /2$. The absorbing planes are $y=\pm y_0$
with $y_0>\pi /2.$ In this setup the motion of the particles in the $x$
direction is independent of that in the $y$ direction. The latter is the
object of the proposed experiment.

Particles that hit the planes $y=\pm y_0$ leave traces at points $%
x_1,x_2,...,x_N$. These distances are proportional (within the constraints
of uncertainty) to the times of arrival at the absorbing walls of particles
that start out in the interval $-\pi /2<y<\pi /2$ with initial velocities
distributed as in eq.(\ref{velocityd}). The histogram obtained from these
points, on an axis normalized with the velocity $v_x,$ is that of the times
of arrival of one dimensional particles moving on the $y$ axis.

The wave function for this configuration is given by 
\[
\psi \left( x,y,t\right) =\psi ^1(x,t)\psi ^2\left( y,t\right) , 
\]
where $\left| \partial \psi ^1(0,t)/\partial x\right| ^2$ is the same as
given in eq.(\ref{dK0t2}) and 
\[
\psi ^2\left( y,t\right) =\sum_{n=1}^\infty \frac 2{n\pi ^{3/2}}\cos \frac{%
n\pi ^2}{4y_0}\sin \frac{n\pi }{y_0}y\exp \left\{ -\frac i\hbar \frac{n^2\pi
^2}{y_0^2}t\right\} . 
\]

According to our theory, the pattern on each lateral wall is given by 
\begin{eqnarray}
{\cal J}(x,\pm y_0) &=&\int_0^\infty {\cal J}(x,\pm y_0,t)\,dt  \label{Jxy0}
\\
&=&\int_0^\infty \left| \frac{\partial \psi ^1(x,t)}{\partial x}\right|
^2\left| \psi ^2(\pm y_0,t)\right| ^2\,dt.  \nonumber
\end{eqnarray}

Next, we consider the pattern observed on the wall $x=0$, $-\pi /2<y<\pi /2$%
. The histogram of the traces of the particles on the screen at $x=0$ is
given by 
\begin{equation}
{\cal J}(y)=\int_0^\infty \left| \partial \psi ^1(0,t)/\partial x\right|
^2\left| \sum_{n=1}^\infty \frac 2{n\pi ^{3/2}}\cos \frac{n\pi ^2}{4y_0}\sin 
\frac{n\pi }{y_0}y\exp \left\{ -\frac i\hbar \frac{n^2\pi ^2}{y_0^2}%
t\right\} \right| ^2\,dt.  \label{Jyx0}
\end{equation}
If the velocities in the $x$ direction are concentrated around $v_x$, the
histogram will be approximately 
\begin{eqnarray*}
{\cal J}(y) &\approx &{\cal J}(y,\bar{t})=\left| \partial \psi ^1(0,\bar{t}%
)/\partial x\right| ^2\psi ^2\left( y,\bar{t}\right) \\
&=&\left| \partial \psi ^1(0,\bar{t})/\partial x\right| ^2\left|
\sum_{n=1}^\infty \frac 2{n\pi ^{3/2}}\cos \frac{n\pi ^2}{4y_0}\sin \frac{%
n\pi }{y_0}y\exp \left\{ -\frac i\hbar \frac{n^2\pi ^2}{y_0^2}\bar{t}%
\right\} \right| ^2.
\end{eqnarray*}
This is not the same as the expression obtained in eq.(\ref{Jy}). The
difference is due to the effect of the lateral absorbing boundaries. Thus,
according to our theory, absorbing boundaries cannot be ignored, as 
usually done in quantum mechanics.

If absorption in energy windows is assumed, the function $\psi ^1(x,t)$ is
the same as calculated in Section 7.1 above,  
\[\psi^2\left(y,t\right)=\sum_{n=1}^\infty\frac2{n\pi^{3/2}}\cos
\frac{n\pi^2}{4y_0}\sin\frac{n\pi}{y_0}y\exp\left\{-\frac i\hbar
\frac{n^2\pi^2}{y_0^2}t-\frac{\lambda n^2\pi^2}{y_0^2}t\right\}.\]
The patterns on the absorbing walls are given by eqs.(\ref{Jxy0}) and 
(\ref{Jyx0}). \\
 \newline 
\noindent
{\large {\bf 8. Discussion and summary}}\newline
\renewcommand{\theequation}{8.\arabic{equation}} \setcounter{equation}{0}

We observe that according to eq.(\ref{S1P1}) there is conservation of
probability: the probability of the absorbed Feynman trajectories and that
of the surviving trajectories sum to 1. This is the result of our
postulate that the Feynman integrals over the two classes of trajectories
have disjoint supports, that is, the absorbed trajectories never return to
the interval $[a,b]$. This conservation of probability persists for all
times. This result is different from that obtained in decoherent state
theory \cite{Halliwell}.

The survival probability $S(t)$ is the probability obtained by repeating
the experiment of observing the particle between the absorbing boundaries
at time $t$ and constructing a histogram of the number of times the
particle is observed. This probability is not a quantum mechanical
quantity in the sense that it is not the integral of the squared modulus
of a probability amplitude defined by Schr\"odinger's equation.

The wave function of the particle at time $t$, given that it has not been
absorbed by that time, is the wave function $\psi(x,t)$ defined by
Schr\"odinger's equation inside the given domain with reflecting boundary
conditions. That is, if the absorbing boundary represents a detector, the
wave function of the particle is $\psi(x,t)$ as long as the particle has
not been detected.

The concept of absorption of particles, energy, momentum, and so on, is an
aspect of time irreversible processes in quantum mechanics. Examples of
irreversible processes in quantum mechanics are the notion of collapse of
the wave function that is caused by measurement (this will discuses  at 
 \cite{measurement} ), the decay of a state of a
particle, a particle that enters into a bulk and loses its momentum due to
interactions with the particles of the bulk, and so on. This paper is an
attempt to construct a formalism for the description of such phenomena.

The process of absorption of quantum particles can be illustrated by a
packet fo identical non-interacting particles that hit an absorbing 
surface, for example, a photographic plate. At the moment a particle 
hits the plate it is absorbed in the sense that its wave function no
longer evolves according to the Hamiltonian of the particles that have
not been absorbed so far. Thus the absorbing surface separates the
particles of the packet into two sets, those that have reached the
surface by a given time and those that have not. The interference
between the Feynman integral over the trajectories of the absorbed
particles and that over the trajectories of the surviving ones
vanishes. The probability of the absorbed particles is discounted from 
tha total probability of the packet. This is also the case for any 
other detector that absorbs particles. The surviving trajectories, 
those that have not reached the absorbing boundaries so far, give rise 
to a reflected wave, as if the absorbing boundary were an infinite 
potential wall. This fact becomes apparent not from a solution of a 
wave equation, but rather from the calculation of the Feynman integral 
over a class of restricted trajectories.

Our derivation does not start with a Hamiltonian, but rather with an action
of trajectories in a restricted class. 

Quantum mechanics without absorption is recovered from our formalism when
the absorbing boundaries are moved to infinity or when the absorption
constant $\lambda$ vanishes.

The examples demonstrate the expected phenomenon that particles that reach
the absorbing boundary are partially reflected and partially absorbed. In
either case the decay pattern of the wave function seems to be new.\\

\noindent
{\bf Acknowledgment:} The authors wish to express their gratitude to
Y. Aharonov and L. Horowitz for useful discussions.\\
 
\end{document}